\let\Xdocument\document
\documentclass{iau}
\let\document\Xdocument

\usepackage{amsmath}
\usepackage{graphicx}
\usepackage{multirow}

\begin{document}

\lefttitle{Akiharu Nakagawa \textit{et al}.}
\righttitle{Period-magnitude relation for massive AGB stars and its astronomical applications}

\jnlPage{1}{7}
\jnlDoiYr{2021}
\doival{10.1017/xxxxx}

\aopheadtitle{Proceedings IAU Symposium}
\editors{C. Sterken,  J. Hearnshaw \&  D. Valls-Gabaud, eds.}

\title{Implication of the period-magnitude relation for massive AGB stars and its astronomical applications}

\author{Akiharu Nakagawa$^{1}$, 
             Tomoharu Kurayama$^{2}$,
             Hiroshi Sudou$^{3}$, 
             and Gabor Orosz$^{4}$}
\affiliation{$^{1}$Kagoshima University, 1-21-35 Korimoto, Kagoshima-shi, Kagoshima 890-0065, Japan \\
                 $^{2}$Teikyo University of Science, 2-2-1 Senju-Sakuragi, Adachi-ku, Tokyo 120-0045, Japan \\
                 $^{3}$Gifu University, 1-1 Yanagido, Gifu City, Gifu 501-1193, Japan\\         $^{4}$Joint Institute for VLBI ERIC (JIVE), Oude Hoogeveensedijk 4, 7991 PD, Dwingeloo, Netherlands\\
}

\begin{abstract}
We present astrometric very long baseline interferometry (VLBI) studies of AGB stars. 
To understand the properties and evolution of AGB stars, distances are an important parameter. 
The distribution and kinematics of their circumstellar matter are also revealed with the VLBI method. 
We used the VERA array to observe 22\,GHz H$_2$O masers in various subclasses of AGB stars. 
Parallaxes of the three OH/IR stars NSV17351, OH39.7$+$1.5, IRC$-$30363, and the Mira-type variable star AW~Tau were newly obtained. 
We present the circumstellar distribution and kinematics of H$_2$O masers around NSV17351. 
The absolute magnitudes in mid-infrared bands of OH/IR stars with very long pulsation periods were investigated and a period-magnitude relation in the WISE W3 band, $M_{\mathrm{W3}} = (-7.21\pm1.18)\log P + (9.25\pm3.09)$, was found for the Galactic AGB stars. 
The VLBI is still a powerful tool for parallax measurements of the Galactic AGB stars surrounded by thick dust shells. 
\end{abstract}


\maketitle

\section{Introduction}
\subsection{Evolution of AGB stars}
Asymptotic giant branch (AGB) stars are known to be the final stage of the evolution of stars with an initial mass of 0.8-10\,$M_{\odot}$ (e.g., \citealp{kar14}). 
Considering the shape of the initial mass function (IMF), a significant portion of stars will spend a period of their lifetime as AGB stars. 
They are surrounded by thick circumstellar dust shells and frequently present stellar pulsations. 
Since AGB stars return various elements into interstellar space by stellar winds, they are important objects that contribute to the chemical composition of the universe and galaxy  (e.g., \citealp{hof18}). 
AGB stars exhibit a wide range of pulsation periods, with the shortest periods being around 100 days and occasionally reaching 3000 days (e.g., \citealp{hab96}). 
Hence, they are often referred to as long-period variable stars (LPVs). 
A more detailed picture of the evolutionary process on the AGB phase reveals several stages depending on their growth. 

During the early AGB phase, stars have relatively thin dust layers, making them observable in both optical and infrared bands. 
The Mira-type variables, visible in both optical and infrared bands, are a well-known class of AGB stars. 
As they progress, strong absorption caused by thick circumstellar dust shells makes them faint and difficult to be detected in optical bands (e.g., \citealp{hab96, kam20}).
On the other hand, they become brighter in the infrared bands as a result of re-radiation from the outer dust layer. 
At this stage, many objects also present OH maser emission from their outermost layer, therefore they are referred to as OH/IR stars. 
The OH/IR stars, a sub-class of AGB stars, are considered to be in the late stage of the AGB phases before progressing to become planetary nebulae \citep{tel91}. 
Additionally, there are other sub-classes that exist in the late stages of the AGB phase. 
Massive stars with initial masses ranging from 7 to 10\,$M_{\odot}$ are known to go through a phase referred to as the super AGB phase, characterized by high luminosity and long pulsation periods \citep{kar19}. 
OH/IR stars with intermediate mass and large mass loss rates ($>$10$^{-4}$\,$M_{\odot}$/yr) are classified as extreme-OH/IR stars \citep{jus15}. 

Mira-type variables, surrounded by thin dust shells, represent the early stage of the AGB phase and are often associated with SiO and H$_2$O masers. 
Following the Mira phase, H$_2$O molecules are transported to the outer side of the circumstellar envelope and are photodissociated, resulting in the production of OH maser emissions (e.g., \citealp{gol17}). 
Due to an excess of infrared emission from the circumstellar dust shell and the presence of OH maser emission, they will be recognized as OH/IR stars (e.g., \citealp{nym98}). 
As they progress towards the post-AGB phase, the amplitude of their stellar pulsation gradually decreases, eventually leading to a phase known as non-variable OH/IR stars (e.g., \citealp{hab96, kam20}). 
Subsequently, they move into the post-AGB phase. 
To understand the sequential evolution from early to late stages in the AGB phases, it is crucial to conduct studies on AGB stars with various properties. 
Target sources of our very long baseline interferometry (VLBI) studies cover a variety of masses and pulsation periods. 
This also means that the target sources that we are studying are representative of a wide range of ages. 

\subsection{Masses of AGB stars and pulsation periods}
There is a correlation between pulsation periods and masses in AGB stars. 
AGB stars with longer pulsation periods are generally considered to be more massive than those with shorter pulsation periods. 
For instance, \cite{fea09} suggests that AGB stars with pulsation periods of 1000 days have masses ranging from 3\,$M_{\odot}$ to 4\,$M_{\odot}$.  
The pulsation periods and mean densities of pulsating stars are known to be coupled (e.g., \citealp{cox80}), and \citet{tak13} also reported a correlation between masses and pulsation periods of AGB stars. 
Therefore, to study AGB stars with various masses, observations of AGB stars exhibiting a wide range of pulsation periods are necessary.

In our previous VLBI observations conducted with the VLBI Exploration of Radio Astrometry (VERA) array from 2003 to 2017, we have carried out many observations towards dozens of AGB stars. 
Most of the target sources are classified as Mira-type variables and SR variables, with pulsation periods typically shorter than 400 days. 
Using H$_2$O masers at 22\,GHz, we have investigated the structures and kinematics of circumstellar matter around Mira and SR variables. 
The study of the SR variable S~Crt by \citet{nak08} was the initial outcome of our VLBI observation program, revealing the parallax and anisotropic circumstellar distribution of H$_2$O masers, as well as detailed investigations of other Mira-type variables. 
Subsequent studies on the Mira-type variables, such as SY~Scl by \citet{nyu11}, R~UMa by \citet{nak18}, and BX~Cam by \citet{mat20}, have also been produced from our VLBI observations. 
However, to date, there have been a limited number of VLBI studies focusing on OH/IR stars. 
Notable examples are the research on U~Her by \cite{vle07} and WX~Psc and OH138.0$+$7.2 by \cite{oro17}, which represent valuable contributions to the OH/IR star studies using the VLBI method.

To obtain a diverse sample of AGB stars covering various evolutionary phases and a range of masses in our observations, it is crucial to include a wide range of pulsation periods. 
Since 2017, we have expanded our VLBI observations to include OH/IR stars with longer pulsation periods. 
The primary objective of our long-term VLBI study is to obtain astrometric and physical properties of OH/IR stars, including distance determination by parallax measurements, proper motion analysis, internal maser motion, luminosity estimation, mass loss rate analysis, and more. 
We think that comparing these properties between OH/IR stars and Mira-type variables can provide insights into their evolutionary relationship. 
In this proceedings paper, we present the current status and results from our astrometric VLBI observations. 
In particular, we will discuss the latest research results on the OH/IR star NSV17351 and present preliminary parallax values for other OH/IR stars.

\subsection{Finding a new period-magnitude relation of Galactic AGB stars}
It is known that there is a relation between K-band apparent magnitudes ($m_\mathrm{K}$) and the logarithm of pulsation periods ($\log P$) of Mira-type variables in the Large Magellanic Cloud (LMC)  (e.g., \citealp{woo99, ita04-1}). 
By utilizing the known distance to the LMC, this relation can be converted into a relation between absolute magnitude in the K band ($M_{\mathrm{K}}$) and $\log P$, thus it can be used as a distance estimator for variable stars. 
However, considering the difference of metallicity between the LMC and our Galaxy, it is crucial to establish a period-magnitude relation using sources in our galaxy. 
To convert apparent magnitudes to absolute magnitudes, parallax distances of the Galactic variable stars are required. 
Over the past decade, we have conducted astrometric VLBI observations for Mira and SR variables, and reported their period-magnitude relation as $M_\mathrm{K} = -3.52 \log P + (1.09 \pm 0.14)$ \citep{nak16}. 
As mentioned in the previous subsection, our previous observations had a limited period coverage of shorter than approximately 400 days.

Compared to typical Mira-type variables, OH/IR stars tend to exhibit longer pulsation periods, sometimes exceeding 1000 days. 
Since the OH/IR stars are surrounded by thick circumstellar dust shells, there is a large amount of extinction in optical bands. 
Sometimes the extinction effects extend into the near-infrared region, including the K band. 
We think this effect leads to a scattering of estimated K-band absolute magnitudes \citep{nak18}. 
Conversely, the extinction diminishes at longer wavelengths, and the re-radiation effect from the dust shell becomes more dominant. 
To mitigate the impact of circumstellar extinction, we aim to validate the existence of the period-magnitude relation in the mid-infrared region. 
We plan to utilize data from the Wide-field Infrared Survey Explorer (WISE)\footnote[1]{\url{http://wise.ssl.berkeley.edu/index.html}}, especially in the W3 band ($\lambda = 12\,\mu$m). 
If a period-magnitude relation is confirmed in the mid-infrared bands, it can serve as a new distance estimator for sources with pulsation periods longer than those typically observed in the Mira-type variables. 
With our new target sources, given in Section~\ref{subsec_tgtsrc}, we will explore this relation in the longer pulsation period range. 

\subsection{Astrometry of OH/IR stars to study the Galactic dynamics}
In recent studies of spiral arms in disk galaxies, there has been a long-standing question about how spiral arms are formed and maintained. 
The quasi-stationary density wave theory (e.g., \citealp{lin64}) and the dynamic spiral theory (e.g., \citealp{swe84, bab15}) are two major theories under discussion.
The studies of the Galactic spiral arms based on three-dimensional $N$-body simulations support a picture of non-steady spiral arms \citep{bab13}. 
The spiral arms do not show rigid rotating patterns but rather differentially rotating dynamic patterns. 
In the dynamic spiral theory, the amplitudes, pitch angles, and pattern speeds of the spiral arms are not constant, but change within a time span of 1-2 rotation periods at each radius \citep{bab15}. 
Characteristic behavior predicted from recent studies is the bifurcating or merging of the Galactic spiral arms on a time scale of $\sim10^8$ yr (e.g., \citealp{bab15}). 
In the Milky Way galaxy, rotation periods at the position of the Sun also correspond to the time scale of $\sim$10$^8$ years.

The OH/IR stars with the longer periods are assumed to have larger masses, i.e. variable stars with $P \simeq 1000$ days have initial masses of 3 to 4\,$M_{\odot}$ \citep{fea09}. 
Assuming this mass range and the relation between the main sequence lifetime $\tau_{\mathrm{MS}}$ and mass given in \citet{spa00}, we obtained $\tau_{\mathrm{MS}}$ of 1.6$\times$10$^8$ to 3.5$\times$10$^8$ years. 
A recent study by \citet{nik22} also supports this estimation. 
Now, we find that the age of OH/IR stars with very long pulsation period ($P \gtrsim 1000$ days) is similar to the characteristic time scale of $\sim10^8$ yr in the dynamic spiral theory. 
This consideration also implies that the age of OH/IR stars with pulsation periods longer than $\sim$1000 days is $\sim$10$^8$ years, which is two orders of magnitude larger than the typical age of high-mass star-forming regions (SFRs) associated with spiral arms. 

In the last 20 years, VLBI astrometry has measured more than two hundred parallaxes of SFRs (e.g., \citealp{bur16, mot16, rei19, ver20}) and evolved stars (e.g., \citealp{kam16a, nak16, nak18, nak19, sud19}). 
However, the ages of almost all VLBI targets fall mainly into two time scales, on the order of 10$^6$ years for the SFRs and 10$^9$ years for the Mira-type variable stars. 
To fully understand the mechanism of spiral arm formation, observations of sources of different ages are now needed, and the very long-period OH/IR stars with estimated ages of $\sim$10$^8$ years can be good probes to fill the time scale gap between 10$^6$ years and 10$^9$ years (Table~\ref{table_corresponding}). 
Astrometric VLBI is a promising tool to determine the three-dimensional positions and kinematics of the OH/IR stars. 
The OH/IR stars that we have selected, which have pulsation periods $P \gtrsim 1000$ days, can contribute to this study.\\

\begin{table}[t!]
\caption{Models and observations for study of the Galactic dynamics.}
\label{table_corresponding}
\begin{center}
\begin{tabular}{cccc} 
    \midrule
Time scale   & Phenomena and model   & Target source &  VLBI Obs.  \\ 
    \midrule
$\sim 10^6$ yr & Spiral arm   & SFRs, giants    & Well studied \\ 
$\sim 10^8$ yr & Bifurcating/merging arm & Heavy OH/IR stars  & \textit{\textbf{Few cases}} \\
$\sim 10^9$ yr & Thick disk stars    & Miras     & Well studied \\
    \midrule
\end{tabular} 
\end{center}  
\end{table} 

\subsection{Why are VLBI observations important?}
Annual parallaxes can be used to derive distances of celestial sources without making any assumptions about their chemical and/or physical properties. 
Recently, {\em Gaia} Data Release\,3\footnote[1]{\url{https://www.cosmos.esa.int/web/gaia/data-release-3}} (DR3; \citealp{gai23}) has provided a huge amount of astrometric measurements. 
Most of the VLBI parallax measurements made so far have been towards SFRs. 
They are very close to the Galactic plane and are deeply obscured by dense dust and molecular clouds. 
As a result, the optical emission from the stars is intercepted by their surroundings and interstellar matter.
For this reason, it is relatively difficult to find counterparts to SFRs in {\em Gaia} catalogs. 
Compared to the SFRs, it is easier to identify AGB stars in {\em Gaia} catalogs because there are many AGB stars distributed at high Galactic latitude. 
Because they are distributed far from the Galactic plane, they are not as heavily obscured by interstellar dust or molecular clouds. 
Mira-type variables, which are thought to be in the early stages of AGB phases, are bright in both optical and infrared bands. 
Parallaxes for a large number of Mira-type variables are available in {\em Gaia} catalogs. 
Since the AGB stars have both {\em Gaia} and VLBI parallax values, they are suitable sources for verifying {\em Gaia} and the VLBI parallax measurements. 

After the release of {\em Gaia} Data Release\,2\footnote[2]{\url{https://www.cosmos.esa.int/web/gaia/dr2}} (DR2), \citet{chi18} conducted a study using three-dimensional radiative hydrodynamics simulations of convection to explore the impact of convection-related surface structures in AGB stars on their photometric variability.
They extracted parallax errors in DR2 for SR variables in the solar neighbourhood and compared them with synthetic predictions of photocenter displacements. 
As a result, they reported that the position of the photocenter has a temporal excursion between 0.077 -- 0.198\,au (5 to 11\% of the corresponding stellar radius), depending on the simulation considered. 
Since the distances of the sources in our VLBI studies are in the order of a hundred pc to a few kpc, the angular size of the excursion in \citet{chi18} can be expected to be 0.1 -- 1\,mas. 
In addition, the time variation of the surface brightness degrades the accuracy of parallax measurements on optical images. 
Therefore, in DR2, and even in its updated version DR3, {\em Gaia}'s parallax measurements of AGB stars can be expected to suffer from this effect. 

If the central star is surrounded by thick dust layers in the late stage of the AGB phase, the source becomes very faint in the optical bands and cannot be observed with Gaia. 
For example, an OH/IR star, OH127.8$+$0.0, is known to have a thick circumstellar dust shell, a high mass loss rate \citep{kem02}, and a long pulsation period of 1994 days (VSX\footnote[1]{\url{https://www.aavso.org/vsx/}}; \citealp{wat06}). 
But we cannot find it in DR3 catalog. 
OH127.8$+$0.0 is very bright in the infrared, but faint in the optical bands due to the strong influence of circumstellar extinction by the dust layer. 
The VLBI method is still a very effective and promising tool for making parallax measurements of this kind of stars.

\section{Observation}
\subsection{Single dish monitoring of H$_2$O and SiO masers}
\label{subsec_single}
Using the 20 m aperture telescope at the VERA Iriki station, we have been observing H$_2$O and SiO maser emissions to obtain their spectra and time variability. 
Since the pulsation periods of a large number of dusty OH/IR stars are not found in the literature or in databases, we have to determine the pulsation period ourselves from single-dish observations.

The integration time of our single-dish observations is 10 to 40 minutes, to reduce the noise level (antenna temperature in K) in each observation to less than 0.05 K. 
The time interval of single-dish observations is approximately one month. 
The conversion factor from antenna temperature to flux density is 19.6 Jy\,K$^{-1}$. 
The acquired signal with a bandwidth of 32 MHz is split into 1024 spectral channels with a frequency resolution of 31.25 kHz, which corresponds to a velocity resolution of 0.42 km\,s$^{-1}$ at 22\,GHz and 0.21 km\,s$^{-1}$ at 43\,GHz. 
We used a signal-to-noise ratio (S/N) of 3 to 5 as a detection criterion in our single-dish observations. 

To measure the overall activity of the circumstellar masers, we use the integrated intensities $I$ in unit of K\,km\,s$^{-1}$ as an integration of the total maser components over a detected velocity range. 
We use the value $I$ to estimate the pulsation period. 
Antenna temperatures have relative uncertainties of 5-20\%, and we have uniformly applied uncertainties of 10\% to all the integrated intensities. 
A simple sinusoidal function $I_{\mathrm{model}}$, defined as 
\begin{eqnarray}
{I_{\mathrm{model}} = {\Delta}I \sin \left[\frac{2\pi(t+\theta)}{P}\right] + I_0}, 
\end{eqnarray} 
is used to estimate the pulsation period.
${\Delta}I$ is the amplitude of the variation, $t$ is the time, $\theta$ is a zero-phase time lag, $P$ is the period of the variation, and $I_0$ is the average. 

\subsection{VLBI observations}
\label{subsec_obsvlbi}
\begin{figure}[th!]
  \begin{center}
    \includegraphics[width=70mm]{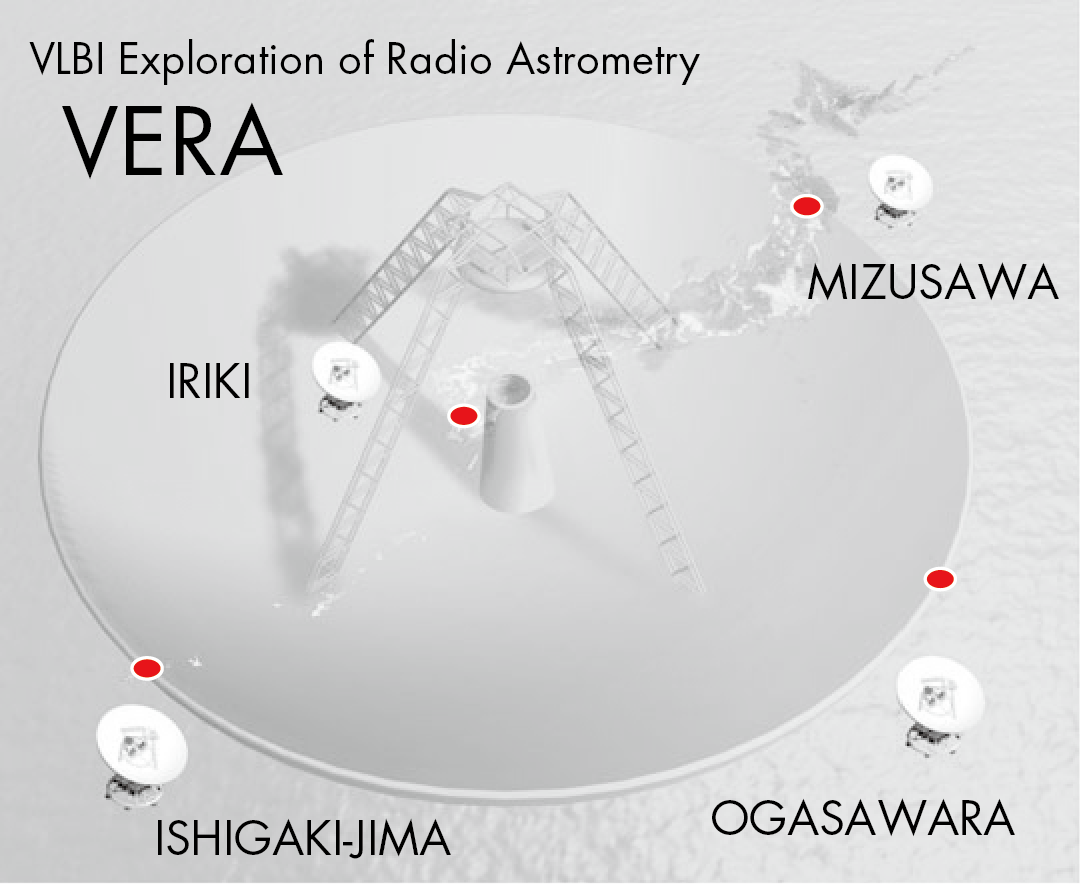}
  \caption{
  Locations of the 4 antennas of the VERA array, which consists of four 20-metre aperture antennas spread across Japan, with the longest baseline length of 2270\,km between Ishigaki-jima and Mizusawa. 
  }
  \end{center}
\label{fig_vera4ant}
\end{figure}

For parallax measurements and mapping of circumstellar masers, we carry out continuous VLBI observations.
VERA, operated by the National Astronomical Observatory of Japan (NAOJ), has been used to observe 22\,GHz H$_2$O and 43\,GHz SiO maser emission from Mira-type variables and OH/IR stars.
The VERA array consists of four 20-metre aperture radio telescopes at Mizusawa, Iriki, Ogasawara, and Ishigaki-jima (Figure~\ref{fig_vera4ant}). 
The array, including the longest baseline length of 2270\,km between Ishigaki-jima and Mizusawa, gives a typical synthesised beam size of $\sim$1.2\,mas and $\sim$0.6\,mas at 22\,GHz and 43\,GHz, respectively. 
Each antenna of the VERA array is equipped with a dual beam system (e.g., \citealp{kaw00}), which allows us to simultaneously observe a target maser source and an extragalactic continuum source at a separation angle between 0.3\,$^{\circ}$ and 2.2\,$^{\circ}$.
The extragalactic sources are used as a position reference. 
Using the dual beam system, we can calibrate short-term tropospheric fluctuations using the phase referencing technique \citep{hon08}.
The relative position of the target maser spots with respect to the position reference source can then be determined with an accuracy of better than 0.1 mas. 
By tracking the celestial motions of the maser spots, the annual parallax of the target is derived. 

The signals of left-handed circular polarization from the target and position reference source are acquired with a total data acquisition rate of 1 giga-bit per second (Gbps). 
This gives a total bandwidth of 256 MHz.
The data were recorded on the hard disks of the ``OCTADISK'' system \citep{oya16}. 
This total bandwidth is divided into 16 intermediate frequency (IF) channels. 
Each IF channel has a width of 16 MHz. 
One IF channel (16 MHz) was allocated to the maser source and the remaining 15 IF channels (16 MHz $\times$ 16 $=$ 240 MHz) were allocated to the position reference sources.
Correlation processing was performed using the Mizusawa software correlator at the Mizusawa VLBI observatory, NAOJ.
In the final output of the correlator, the 16 MHz bandwidth data of the H$_2$O or SiO masers were divided into 512 channels with a frequency resolution of 31.25 kHz.
This corresponds to a velocity resolution of 0.42 km\,s$^{-1}$ at 22\,GHz and 0.21 km\,s$^{-1}$ at 43\,GHz.

For reduction of the VLBI data we use the Astronomical Image Processing System\footnote[1]{\url{http://www.aips.nrao.edu/index.shtml}} (AIPS; \citealp{fom81}) developed by the National Radio Astronomy Observatory (NRAO).
A detailed description of the phase referencing analysis is given in \cite{nak23a}. 

\subsection{Phase referencing with the VERA dual beam system}
\label{subsec_pref}
We will give a simplified explanation of the phase referencing method using the dual beam system equipped with the VERA array. 
Letting the phases obtained by a VLBI observation towards the target maser source A and the reference continuum source B be $\phi_{{\rm obs}}^{{\rm A}}$ and $\phi_{{\rm obs}}^{{\rm B}}$, respectively, we can express the phases as follows:  
\begin{eqnarray}
\phi_{{\rm obs}}^{{\rm A}} = 
\phi_{{\rm pos}}^{{\rm A}} + \phi_{{\rm struc}}^{{\rm A}} + 
\phi_{{\rm atm}}^{{\rm A}} + \phi_{{\rm inst}}^{{\rm A}} + \phi_{{\rm clock}}^{{\rm A}} 
\label{eq_tho01_12.23_A}
\end{eqnarray}
and 
\begin{eqnarray}
\phi_{{\rm obs}}^{{\rm B}} = 
\phi_{{\rm pos}}^{{\rm B}} + \phi_{{\rm struc}}^{{\rm B}} + 
\phi_{{\rm atm}}^{{\rm B}} + \phi_{{\rm inst}}^{{\rm B}} + \phi_{{\rm clock}}^{{\rm B}} , 
\label{eq_tho01_12.23_B}
\end{eqnarray}
where the terms on the right-hand side are 
the residual phase due to the sky plane position of the sources ($\phi_{{\rm pos}}^{{\rm A}}$), 
the residual phase due to the source structure ($\phi_{{\rm struc}}$), 
the residual phase due to the unpredictable atmospheric path length ($\phi_{{\rm atm}}$), 
the residual phase due to the difference between the two signal paths ($\phi_{{\rm inst}}$), 
and the residual phase due to the clock offset ($\phi_{{\rm clock}}$) 
(see, e.g., \citealp{tho01}). 
Since the two sources A and B have a small separation angle on the sky plane, $\phi_{{\rm atm}}^{{\rm A}}$ and $\phi_{{\rm atm}}^{{\rm B}}$ are considered to be equal. 
The clock offsets $\phi_{{\rm clock}}^{{\rm A}}$ and $\phi_{{\rm clock}}^{{\rm B}}$ can be eliminated in the ordinary reduction process. 
However, since the signal paths for sources A and B are independent in the VERA receiver system, $\phi_{{\rm inst}}^{{\rm A}}$ and $\phi_{{\rm inst}}^{{\rm AB}}$ are not the same,  
\begin{eqnarray}
\phi_{{\rm inst}}^{{\rm A}} \neq \phi_{{\rm inst}}^{{\rm B}}. 
\label{instrumental-phi}
\end{eqnarray}
By taking the difference between the phases from sources A and B, we will obtain 
\begin{eqnarray}
\phi_{{\rm {\rm obs}}}^{{\rm A}} - \phi_{{\rm obs}}^{{\rm B}}= 
 \phi_{{\rm pos}}^{{\rm A}} - \phi_{{\rm pos}}^{{\rm B}} 
+\phi_{{\rm inst}}^{{\rm A}} - \phi_{{\rm inst}}^{{\rm B}} .
\label{eq_sak_3.1}
\end{eqnarray}
Here we use the calibration information obtained from the dual beam system.
Artificial noise sources are installed on the surface of the VERA antenna and the same noise is sent to the two receivers along with the signal from the celestial source \citep{hon08}.
By correlating two signals from A and B, the phase error caused by the difference in path length $\phi_{{\rm inst}}^{{\rm A}} - \phi_{{\rm inst}}^{{\rm B}}$ can be obtained. 
The phase $\phi_{{\rm pos}}^{{\rm B}}$ is assumed to be zero for the source with point-like structure. 
Even if source B has some structure, we can estimate and eliminate the $\phi_{{\rm pos}}^{{\rm B}}$ term by the self-calibration reduction process. 
Finally, we obtain 
\begin{eqnarray}
\phi_{{\rm {\rm obs}}}^{{\rm A}} - \phi_{{\rm obs}}^{{\rm B}}= 
 \phi_{{\rm pos}}^{{\rm A}}.
\label{eq_sak_3.1}
\end{eqnarray}
This gives the phase information that reflects the celestial position of the target maser source A. 
By performing VLBI observations for 1.5 year to 2 years with an interval of about one month, we can track the position of the target maser source A to derive its parallax and proper motions. 

\subsection{Target sources}
\label{subsec_tgtsrc}
From 2003 to 2017, we have carried out many VLBI observations of dozens of AGB stars using the VERA array.  
The main targets of the previous studies were Mira-type variables and SR variables, which in most cases have pulsation periods shorter than 400 days \citep{nak18}. 
From 2017, we started VLBI observations of OH/IR stars or Mira-type variables, which have longer pulsation periods than those in the previous studies. 
Among them, the OH/IR stars are shown in particular in Table~\ref{table-src}.  
The first four columns show the sources for which observations have been made between the years 2017 and 2022.
Although successful observations of all sources have been difficult, data acquisition for some stars has been completed and data reduction is currently underway. 
In this proceedings paper, we will present some parallax measurements as preliminary results. 
In early 2023 we proposed VLBI observations of new sources, which are presented in the last four columns of Table~\ref{table-src}. 
The parallaxes from DR3 ($\Pi_{\mathrm{DR3}}$) and their relative errors are also shown. 
For those sources for which no parallax values were found in DR3, we have indicated ``n/a" in their respective ``$\Pi_{\mathrm{DR3}}$" and ``Err." columns. 
For some sources (RAFGL\,5201, RAFGL\,2445, and OH 358.23$+$0.11), the parallaxes have negative values. 
Several sources have errors greater than 100\,\% (NSV17351, RAFGL\,5201, and OH 358.23$+$0.11). 
From this table, we can interpret that it is very difficult for {\em Gaia} to determine accurate parallaxes of OH/IR stars. 
Therefore, the VLBI method is still important for parallax measurements of these dust-obscured OH/IR stars. 
Pulsation periods are also given in columns 4 and 8.
We have selected sources that cover a wide range in pulsation periods. 
Sources at low Galactic latitudes with longer pulsation periods can be expected to be young and more massive than typical Mira-type variables. 

\begin{table}[th!]
 \centering
 \caption{DR3 parallaxes of our VLBI target sources}
 \label{table-src}
 {\tablefont\begin{tabular}{@{\extracolsep{\fill}}lcccclccc}
    \midrule
Source name & $\Pi_{\mathrm{DR3}}$ & Err. & Period & & 
Source name & $\Pi_{\mathrm{DR3}}$ & Err. & Period\\
(2017--2022)$^{\dag}$ & [mas] & [$\%$] & [days] & & 
(2023--)$^{\dag}$     & [mas] & [$\%$] & [days]\\
    \midrule
NSV17351 & 0.088$\pm$0.147 & 166 & 1122       &  & V697\,Her & 1.029$\pm$0.129 & 13 & 497\\
OH\,127.8$-$0.0 & n/a  & n/a & 1994           &  & NSV\,23099 & 0.209$\pm$0.102 & 49 & 431\\
NSV\,25875 & n/a & n/a  & 1535                 & & OH\,358.667$-$0.044 & 0.207$\pm$0.142 & 69 & 300 \\
RAFGL\,5201 & $-$0.131$\pm$0.253 & $-$194 & 600  & & OH\,358.23$+$0.11 & $-$0.061$\pm$0.190 & $-$313 & 704\\
OH\,83.4$-$0.9 & 0.836$\pm$0.556 & 66     &1428  & & OH\,0.66$-$0.07 & n/a & n/a & n/a\\
OH\,141.7$+$3.5 & n/a & n/a                 &1750  & & IRAS\,18039$-$1903 & n/a & n/a & n/a\\
CU\,Cep & 0.231$\pm$0.057 & 25            & 700 & &  OH\,9.097$-$0.392 & 0.261$\pm$0.232 & 89 & 634\\
RAFGL\,2445 & $-$1.548$\pm$0.369 & $-$24  & n/a  & &  RAFGL\,1686 & 1.053$\pm$0.359 & 34 & 500\\
OH\,39.7$+$1.5 & n/a & n/a            & 1260 & &  IRAS\,18176$-$1848 & 2.404$\pm$0.618 & 26 & n/a\\
OH\,26.5$+$0.6 & n/a & n/a            & 1589 &  & OH\,44.8$-$2.3 & 0.918$\pm$0.631 & 69 & n/a\\
OH\,42.3$-$0.1 & n/a & n/a            & 1650 &  &   &  &  & \\
IRC$-$30363 & 0.241$\pm$0.130 & 54       &  720 &  &   &  &  & \\
IRC$+$10322 & 0.553$\pm$0.183 & 33       &  570 &    & &  &  & \\
IRC$+$10451 & 0.818$\pm$0.196 & 24       &  730 &    & &  &  & \\
OH\,26.2$-$0.6 & n/a & n/a            & 1330 & &    &  &  & \\
OH\,51.8$-$0.1 & n/a & n/a            & 1270 &   &  &  &  & \\
OH\,358.16$+$0.49 & n/a & n/a        & 1507  &  &    &  & \\
V1018\,Sco & n/a & n/a               & n/a &  &    &  & \\
    \midrule
    \end{tabular}}
    \tabnote{${\dag}$ Duration of our VLBI observations using VERA.}
\end{table}

\section{Results and discussion}
\subsection{Parallaxes from VLBI and {\em Gaia}}
\label{sub_parallax}
We will now compare parallax measurements of AGB stars from VLBI and {\em Gaia}. 
In Table~\ref{table-pi} we have presented 44 Galactic LPVs whose parallaxes have been determined from astrometric VLBI, in order of their right ascension (RA). 
In the third, fourth, and fifth columns, the parallax values determined from astrometric VLBI, DR2 and DR3 are presented as $\Pi_{\mathrm{VLBI}}$, $\Pi_{\mathrm{DR2}}$, and $\Pi_{\mathrm{DR3}}$, respectively. 
Although the number of digits of the original parallaxes and their formal errors in DR2/DR3 is large, we have presented the values in digits of 0.001 mas. 
The species of the maser molecules observed in each VLBI observation are given in the sixth column. 
In the last column, the references of their VLBI parallaxes are given using abbreviations, whose meaning is explained in the table footnote. 
Regarding R~Aqr, since the VLBI parallax is published in two independent studies by \cite{kam10} and \cite{min14}, it has two lines in the table. 
For the two sources WX~Psc and OH138.0$+$7.2 we were not able to find the corresponding parallaxes in either DR2 or DR3. 
For W~Leo and Y~Lib we could not find their parallaxes in DR2, but found them in DR3. 
It is assumed that the data quality has improved from DR2 to DR3.
In the case of W~Hya, the annual parallax was listed in DR2, but we could not find it in DR3.
The parallax of S~Ser shows a negative value in DR3. 

\begin{table}[th!]
 \centering
 \caption{Parallaxes from the VLBI and DR2, DR3}
 \label{table-pi}
{\tablefont\begin{tabular}{@{\extracolsep{\fill}}lcccccc}
    \midrule
Source & Var. & $\Pi_{\mathrm{VLBI}}$ & $\Pi_{\mathrm{DR2}}$ & $\Pi_{\mathrm{DR3}}$  & Maser   & Ref.$^{\dag}$ \\
       &    type  &   [mas]             &   [mas]        & [mas]           &            & $\Pi_{\mathrm{VLBI}}$ \\ 
    \midrule
SY~Scl &     Mira & 0.75$\pm$0.03    & 0.675$\pm$0.227 & 0.525$\pm$0.122 &H$_2$O& nyu11\\ 
WX~Psc &    OH/IR& 5.3$^{\,b}$        & n/a &  n/a  & OH    &  oro17\\ 
S~Per &         SRc & 0.413$\pm$0.017& 0.222$\pm$0.121 & $-$0.503$\pm$0.081 &H$_2$O& asa10\\
OH138.0$+$7.2&OH/IR&0.52$\pm$0.09& n/a & n/a & OH     & oro17\\ 
V637~Per & Mira & 0.94$\pm$0.02  & 1.846$\pm$0.152  & 0.845$\pm$0.097 & H$_2$O & ver20\\
BX~Eri & Mira & 2.116$\pm$0.105  & 2.477$\pm$0.110  & 2.349$\pm$0.063 & H$_2$O  & ver20\\
T~Lep  & Mira & 3.06$\pm$0.04   & 2.958$\pm$0.189  & 3.086$\pm$0.103  & H$_2$O& nak14\\ 
BW~Cam & Mira & 0.749$\pm$0.189 & 1.187$\pm$0.214  & 0.956$\pm$0.105 & H$_2$O& ver20\\ 
RW~Lep & Mira   & 1.62$\pm$0.16 & 2.355$\pm$0.134 & 2.539$\pm$0.075  &H$_2$O& kam14\\ 
BX~Cam & Mira & 1.73$\pm$0.03  & 4.134$\pm$0.255  & 1.764$\pm$0.101 & H$_2$O  & mat20\\
U~Lyn  & Mira & 1.27$\pm$0.06    & 0.580$\pm$0.224 & 1.014$\pm$0.083 &H$_2$O& kam16a\\ 
NSV17351 & Mira & 0.247$\pm$0.035  & 0.353$\pm$0.228 & 0.088$\pm$0.147 & H$_2$0  & nak23a\\
VY~CMa & SRc & 0.88$\pm$0.08  &$-$5.917$\pm$0.825 &  0.419$\pm$0.408  &H$_2$O& cho08\\
OZ~Gem & Mira & 0.806$\pm$0.039  &$-$0.961$\pm$0.456 & 0.458$\pm$0.325 &H$_2$O& ura20\\
OH231.8+4.2 &OH/IR& 0.61$\pm$0.03&  0.096$\pm$0.182 & 0.030$\pm$0.160 & H$_2$O& nak23b\\ 
HU~Pup & Mira & 0.308$\pm$0.042  & 0.182$\pm$0.057  & 0.294$\pm$0.030 & H$_2$O & ver20\\
R~Cnc & Mira & 3.84$\pm$0.29   & 4.435$\pm$0.549  & 3.938$\pm$0.179  &H$_2$O&ver20\\ 
X~Hya & Mira & 2.07$\pm$0.05  & 1.891$\pm$0.276  & 2.531$\pm$0.111 & H$_2$O & ver20\\
R~UMa  & Mira & 1.97$\pm$0.05  & 2.045$\pm$0.202 & 1.747$\pm$0.086 &H$_2$O& nak16\\ 
W~Leo & Mira & 1.03$\pm$0.02  &  n/a  & 0.878$\pm$0.108 & H$_2$O & ver20\\
HS~UMa & Mira & 2.816$\pm$0.095  & 3.215$\pm$0.144  & 3.202$\pm$0.101 & H$_2$O & ver20\\
S~Crt  & SRb  & 2.33$\pm$0.13   & 2.646$\pm$0.146 & 0.061$\pm$0.097 &H$_2$O& nak08\\ 
T~UMa  & Mira &   0.96$\pm$0.15 & 0.748$\pm$0.105 & 0.989$\pm$0.065 & H$_2$O& nak18\\ 
U~CVn & Mira & 0.911$\pm$0.031  & 0.921$\pm$0.167  & 0.563$\pm$0.077 & H$_2$O  & ver20\\
RT~Vir & SRb& 4.417$\pm$0.134  & 2.050$\pm$0.291 & 4.137$\pm$0.227 &H$_2$O& zha17\\
R~Hya  & Mira & 7.93$\pm$0.18   & 4.468$\pm$0.894 & 6.736$\pm$0.464 &H$_2$O& ver20\\ 
W~Hya  & Mira &10.18$\pm$2.36  & 6.091$\pm$0.816  & n/a &OH    & vle03\\
RX~Boo & SRb  & 7.31$\pm$0.50   & 7.829$\pm$0.300  & 6.424$\pm$0.231 &H$_2$O& kam12\\ 
FV~Boo& Mira  & 0.97$\pm$0.06   & 0.573$\pm$0.181  &  1.014$\pm$0.091 & H$_2$O& kam16b\\ 
Y~Lib  & Mira &   0.855$\pm$0.050   & n/a  & 0.832$\pm$0.083  &  H$_2$O&chi19\\ 
S~CrB  & Mira & 2.39$\pm$0.17   & 2.322$\pm$0.285 & 2.596$\pm$0.114 &  OH    & vle07\\
S~Ser & Mira & 1.25$\pm$0.04  & $-$0.512$\pm$0.317 & 0.768$\pm$0.129 & H$_2$O & ver20\\
U~Her  & Mira & 3.76$\pm$0.27   &1.749$\pm$0.149 & 2.357$\pm$0.077 &OH    & vle07\\
VX~Sgr & SRc&0.64$\pm$0.04   & 0.787$\pm$0.229 & 0.050$\pm$0.187 &  H$_2$O  & xu18\\
RR~Aql & Mira & 1.58$\pm$0.40   & 3.146$\pm$0.298  & 1.953$\pm$0.113 &  OH    & vle07\\
SY~Aql & Mira & 1.10$\pm$0.07   & 3.433$\pm$0.206 & 1.067$\pm$0.091 &H$_2$O&ver20\\ 
NML~Cyg&SRc&0.62$\pm$0.047& 1.526$\pm$0.568 & 0.528$\pm$0.348 & H$_2$O & zha12\\
UX~Cyg & Mira & 0.54$\pm$0.06 & 0.176$\pm$0.167 & 0.701$\pm$0.094 &  H$_2$O& kur05\\
SV~Peg& SRb& 3.00$\pm$0.06  & 1.124$\pm$0.283  & 2.586$\pm$0.170 &H$_2$O  & sud19\\
IRAS22480$+$6002&SRc&0.400$\pm$0.025&0.479$\pm$0.078& 0.363$\pm$0.029 &H$_2$O& ima12\\
R~Peg  & Mira & 2.76$\pm$0.28   & 2.830$\pm$0.254  & 2.629$\pm$0.117 & H$_2$O&ver20\\ 
R~Aqr  & Mira & 4.7$\pm$0.8       & 3.122$\pm$0.278  & 2.593$\pm$0.333  &SiO   & kam10\\
R~Aqr  & Mira & 4.59$\pm$0.24    & 3.122$\pm$0.278  & 2.593$\pm$0.333 & SiO   & min14\\
PZ~Cas & SRc &  0.356$\pm$0.026  & 0.420$\pm$0.081 & 0.355$\pm$0.041  & H$_2$O& kus13\\
R~Cas  & Mira & 5.67$\pm$1.95  & 5.342$\pm$0.245  & 5.742$\pm$0.201 & OH    & vle03\\
    \midrule
    \end{tabular}}
    \tabnote{${\dag}$ 
Reference of VLBI parallax :  
(nyu11) \cite{nyu11},
(oro17) \cite{oro17},
(asa10) \cite{asa10},
(ver20) \cite{ver20},
(nak14) \cite{nak14},
(kam14) \cite{kam14},
(mat20) \cite{mat20},
(kam16a) \cite{kam16a},
(nak23a) \cite{nak23a},
(cho08) \cite{cho08},
(ura20) \cite{ura20},
(nak23b) This work, preliminary,
(nak16) \cite{nak16},
(nak08) \cite{nak08},
(nak18) \cite{nak18},
(zha17) \cite{zha17},
(vle03) \cite{vle03},
(kam12) \cite{kam12},
(kam16b) \cite{kam16b},
(chi19) \cite{chi19},
(vle07) \cite{vle07},
(xu18) \cite{xu18},
(zha12) \cite{zha12},
(kur05) \cite{kur05},
(sud19) \cite{sud19},
(ima12) \cite{ima12},
(kam10) \cite{kam10},
(min14) \cite{min14},
(kus13) \cite{kus13}.}
\end{table}

\begin{figure}[th!]
  \begin{center}
    \includegraphics[width=130mm]{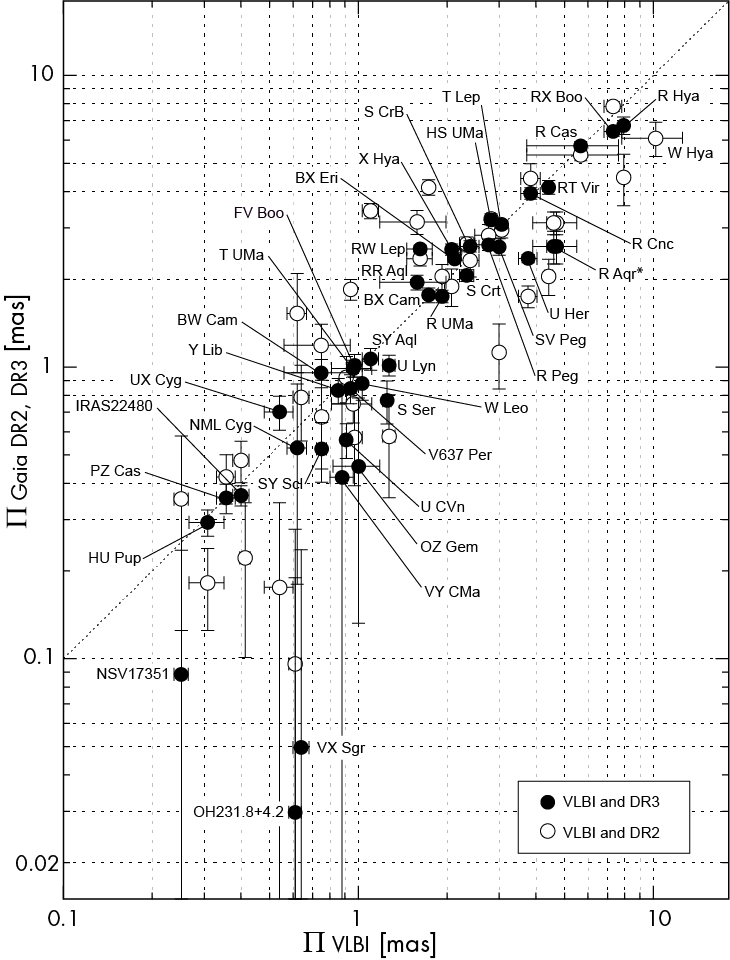}
  \caption{
  Annual parallaxes of 41 LPVs determined from the VLBI (horizontal axis) and DR2/DR3 (vertical axis) on a logarithmic scale. 
  Open circles correspond to the comparison between the VLBI and DR2,
  whereas filled circles correspond to the comparison between the VLBI and DR3. 
  Error bars are also shown on both axes. }
  \end{center}
\label{fig-pivlbigaia}
\end{figure}

In Figure~\ref{fig-pivlbigaia} we present the parallaxes of AGB stars determined from VLBI ($\Pi_{\mathrm{VLBI}}$) and DR2/DR3 ($\Pi_{\mathrm{Gaia\,DR2}}$, $\Pi_{\mathrm{Gaia\,DR3}}$). 
A total of 41 sources are included in the figure. 
The horizontal and vertical axes represent the parallax values on logarithmic scales. 
Open circles represent the comparison between the parallaxes from the VLBI and DR2. 
Filled circles represent the comparison between the VLBI and DR3. 
A dotted line shows a relation of the form $\Pi_{\mathrm{VLBI}} = \Pi_{\mathrm{Gaia\,DR2/DR3}}$. 
In Figure~\ref{fig-pivlbigaia}, source names are added near the comparison data between the VLBI and DR3 (filled circles). 
In the case of W~Hya, we do not have a DR3 parallax, so the name is found near the open circle. 
The dispersion of the filled circles from the $\Pi_{\mathrm{VLBI}} = \Pi_{\mathrm{Gaia\,DR2/DR3}}$ relation is significantly smaller than that of the open circles. 
This suggests that, for many sources, DR3 parallax measurements are closer to those carried out with the VLBI than are those obtained with DR2. 
However, for some sources there are still large discrepancies between the VLBI and DR3 parallaxes. 
For three sources, NSV17351, OH231.8$+$4.2, and VX~Sgr, DR3 measurements are smaller than the VLBI measurements. 
The types of the three LPVs are OH/IR stars (NSV17351 and OH231.8$+$4.2) and red supergiant (VX~Sgr). 
Although we are curious about many other OH/IR stars in Table~\ref{table-src}, they could not be shown in Figure~\ref{fig-pivlbigaia} because {\em Gaia} and/or VLBI parallaxes of many dust-obscured OH/IR stars are not available.  

\subsection{Parallax and error}
\begin{figure}[th!]
\begin{center}
\includegraphics[width=110mm]{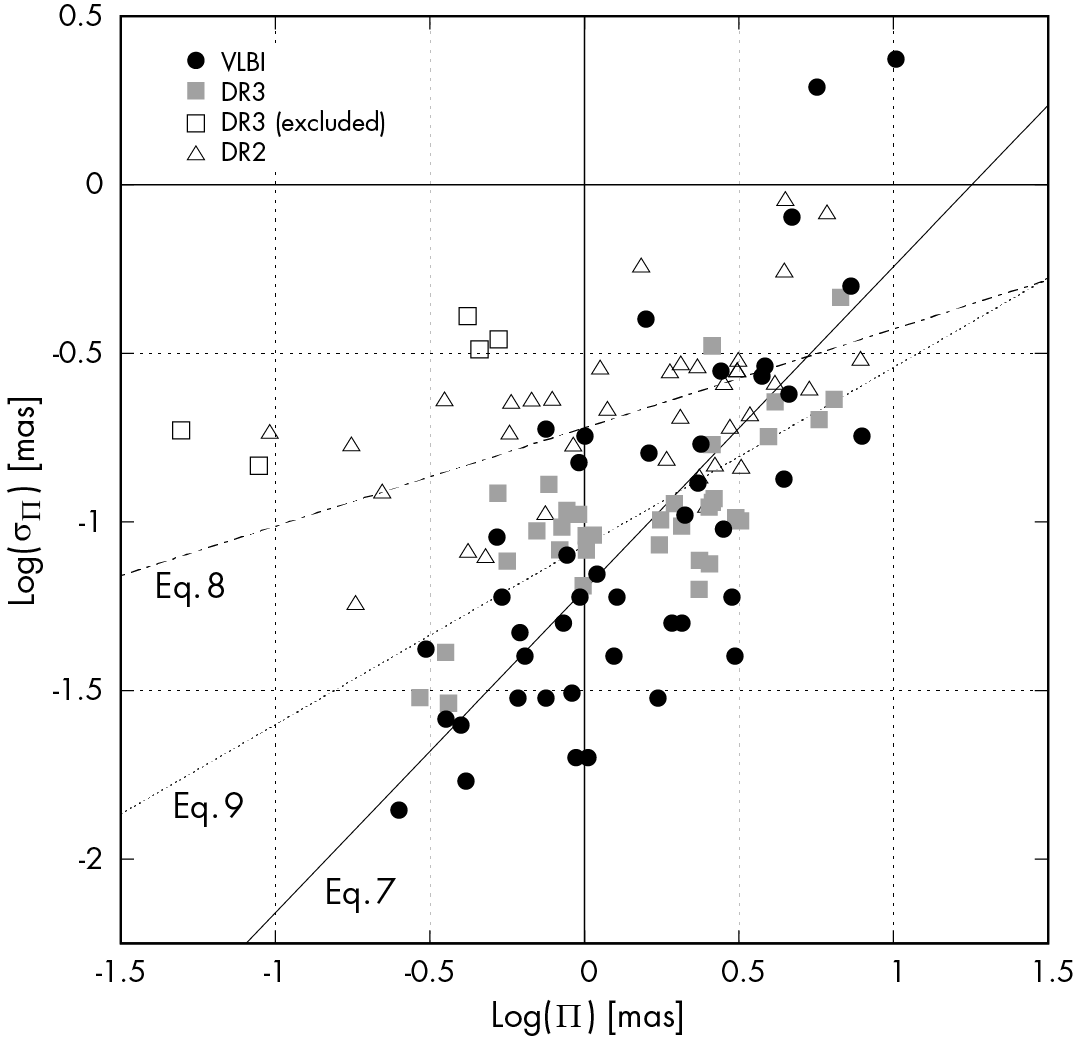}
\end{center}
\caption{
Parallaxes and their formal errors obtained with the VLBI (filled circles), DR2 (triangles), and DR3 (squares). 
Three lines represent the relationships between parallax and errors for the VLBI (solid line; Eq.\,7), DR2 (one-dotted chain line; Eq.\,8), and DR3 (dotted line; Eq.\,9). 
The points of intersection are $\log\Pi = 0.716$ for equations (7) and (8), and $\log\Pi = 0.302$ for equations (7) and (9).
}
\label{fig_pierr}
\end{figure}

We will consider the parallaxes and their errors. 
In Figure~\ref{fig_pierr} we have presented the parallaxes and their formal errors obtained from the VLBI (filled circles), DR2 (open triangles), and DR3 (grey squares). 
From this figure we can see a positive correlation between the parallaxes and the errors for all data sets from the VLBI and DR2/DR3. 
We fitted the distribution to a linear function using a least-squares analysis and obtained three lines as follows, 
\begin{eqnarray}
\label{eq_sak_3.1}
\log\sigma_{\Pi_{\mathrm{VLBI}}}      & = &  (0.96\pm0.13) \log\Pi_{\mathrm{VLBI}} - 1.20 \pm 0.06 \,,\\
\log\sigma_{\Pi_{\mathrm{Gaia\,DR2}}} & = &  (0.29\pm0.07) \log\Pi_{\mathrm{Gaia\,DR2}} - 0.72 \pm 0.04 \,,\\ 
\log\sigma_{\Pi_{\mathrm{Gaia\,DR3}}} & = & (0.53\pm0.09) \log\Pi_{\mathrm{Gaia\,DR3}} - 1.07 \pm 0.03\,.
\end{eqnarray}
It should be noted that five data points from DR3, presented with open squares, were excluded from this consideration because they were distributed outside the main group of data. 
Equations (7) to (9) show the relationships between parallax and errors for the VLBI (solid line), DR2 (one-dotted chain line), and DR3 (dotted line), respectively. 
Equations (7) and (8) intersect at $\log\Pi = 0.716$, corresponding to a distance of 192\,pc. 
Equations (7) and (9)  intersect at $\log\Pi = 0.302$, corresponding to a distance is 499\,pc. 
Using the error ratio $\sigma_{\Pi}/\Pi$ as an indicator of the effectiveness of parallax measurements, VLBI measurements can derive better distance estimates than DR3 for the LPVs further than $\sim499$\,pc. 
We can understand that the VLBI and {\em Gaia} are complementary for distance measurements of dusty AGB stars, and the distance of $\sim$500\,pc would be a boundary of validity between the VLBI and {\em Gaia} for parallax measurement of AGB stars. 

\subsection{Results for individual sources}
\subsubsection{Pulsation period of an OH/IR star NSV17351}
\begin{figure}
\begin{center}
\includegraphics[width=130mm]{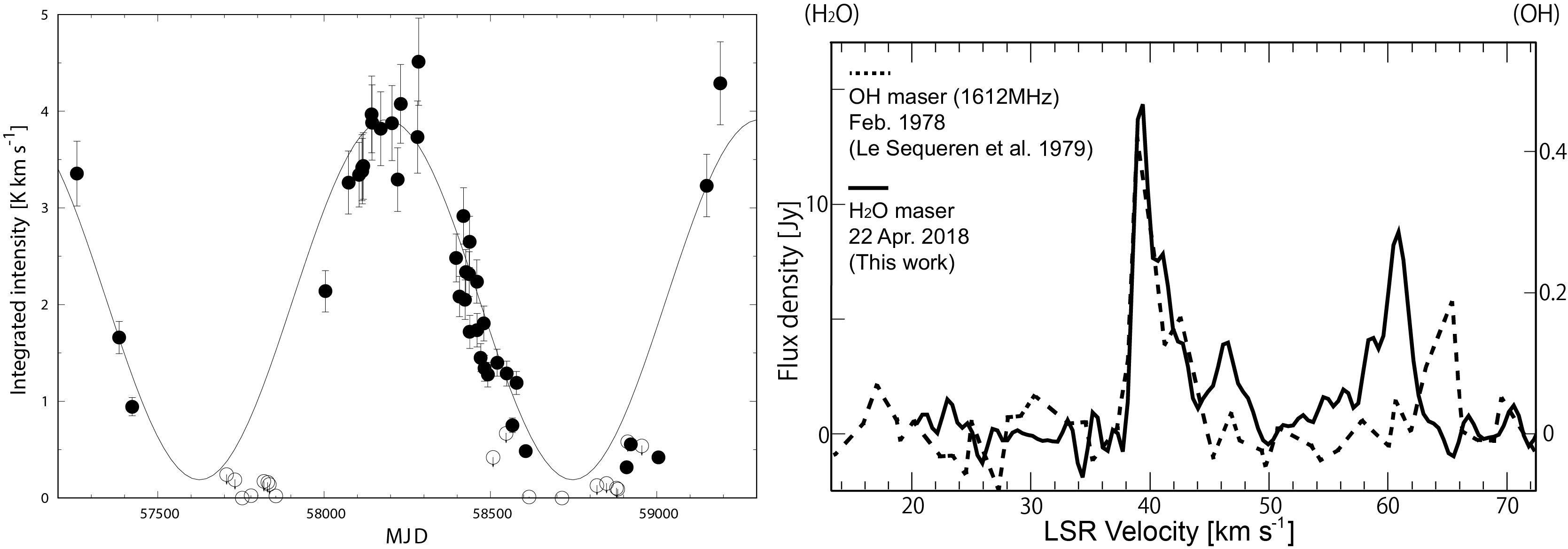}
\caption{
(Left): Time variation of the integrated H$_2$O maser intensities of NSV17351. 
Filled circles represent successful detections. 
In the case of non-detections, open circles with downward arrows represent the upper limits of detection.
The solid line is the model indicating a pulsation period of 1122$\pm$24 days. 
(Right): Superpositions of the H$_2$O maser (solid line) and OH maser (dotted line) of NSV17351 obtained in 2018 and 1978, respectively. 
The cut-off velocity of the blue-shifted side appears to be exactly the same in both spectra. 
}
\label{fig-lightcurve}
\end{center}
\end{figure}

NSV17351 is classified as an OH/IR star~\citep{les79}. 
From long-term monitoring of our single-dish program, we find that this source can be a variable star. 
Although we searched for its pulsation period in online databases and the literature, we could not find it. 
We determined the pulsation period of NSV17351 from our single-dish monitoring of the H$_2$O maser at 22\,GHz. 
From our least-squares analysis assuming a simple sinusoidal function as presented in Section~\ref{subsec_single}, the pulsation period of NSV17351 was solved to be 1122$\pm$24 days \citep{nak23a}. 
The fit of the model is shown with a solid curve in the left panel of Figure~\ref{fig-lightcurve}. 
As there is no prior information on this period in the literature or online databases, this is the first time we have successfully measured its periodicity. 
Since NSV17351 has a very long periodicity, we think that it is a candidate for an extreme OH/IR star. 

In the right panel of Figure~\ref{fig-lightcurve}, we superimposed the H$_2$O maser spectrum obtained on 22 April 2018 (solid line) and a 1612 MHz OH maser spectrum obtained in February 1978 (dotted line). 
We can see that the cut-off velocity in the blue-shifted component shows exactly the same value (38\,km\,s$^{-1}$ to 40\,km\,s$^{-1}$). 
Since it is thought that OH molecules are supplied by photodissociation of H$_2$O molecules carried to the outer part of the circumstellar envelope, we understand that the H$_2$O molecules have been carried to the outermost region and the H$_2$O gas has accelerated to the terminal velocity. 

\subsubsection{Parallax of NSV17351}
\begin{figure}
  \begin{center}
    \includegraphics[width=130mm]{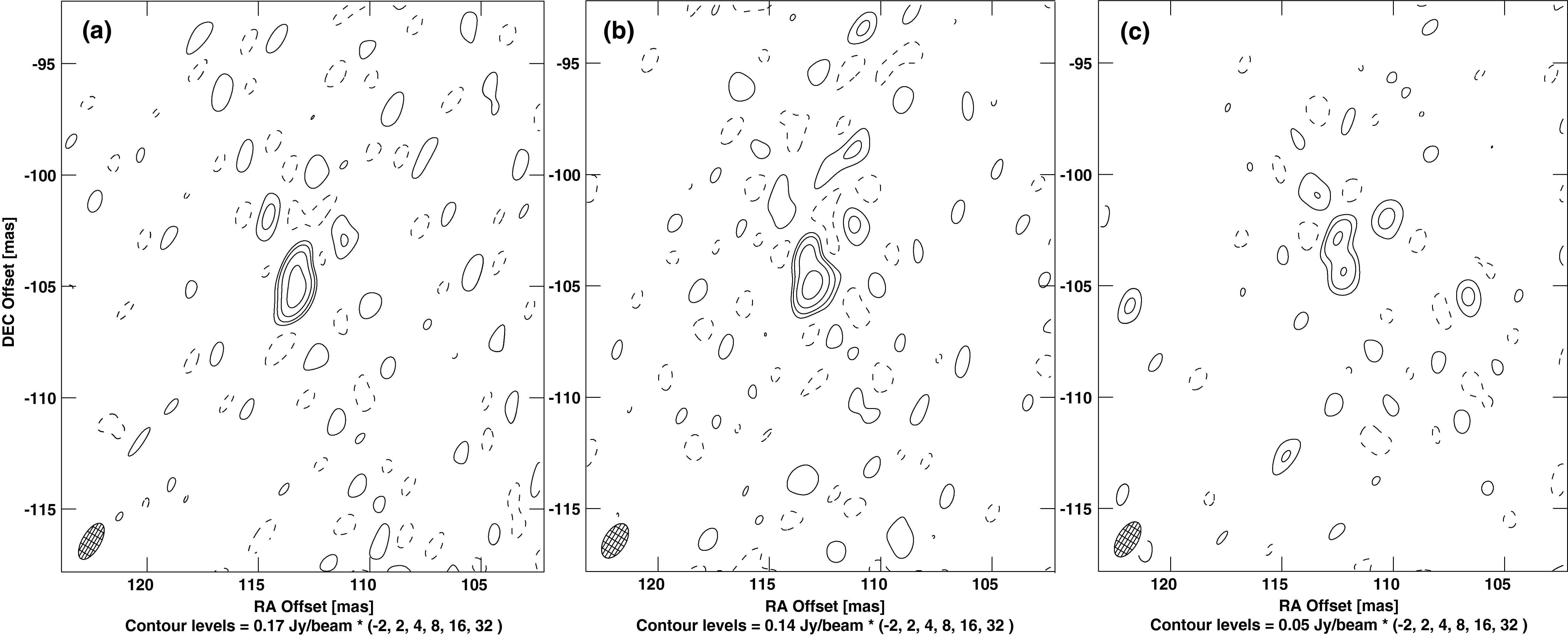}
  \caption{
  VLBI images of maser spots of NSV17351 at a $V_{\mathrm{LSR}}$ of 39.15 km\,s$^{-1}$ detected on (a) 16 April 2018, (b) 1 November 2018 and (c) 12 March 2019 \citep{nak23a}. 
  The synthesised beams are presented in the lower left of each map. }
  \end{center}
\label{fig-3map}
\end{figure}

\begin{figure}
  \begin{center}
    \includegraphics[width=100mm]{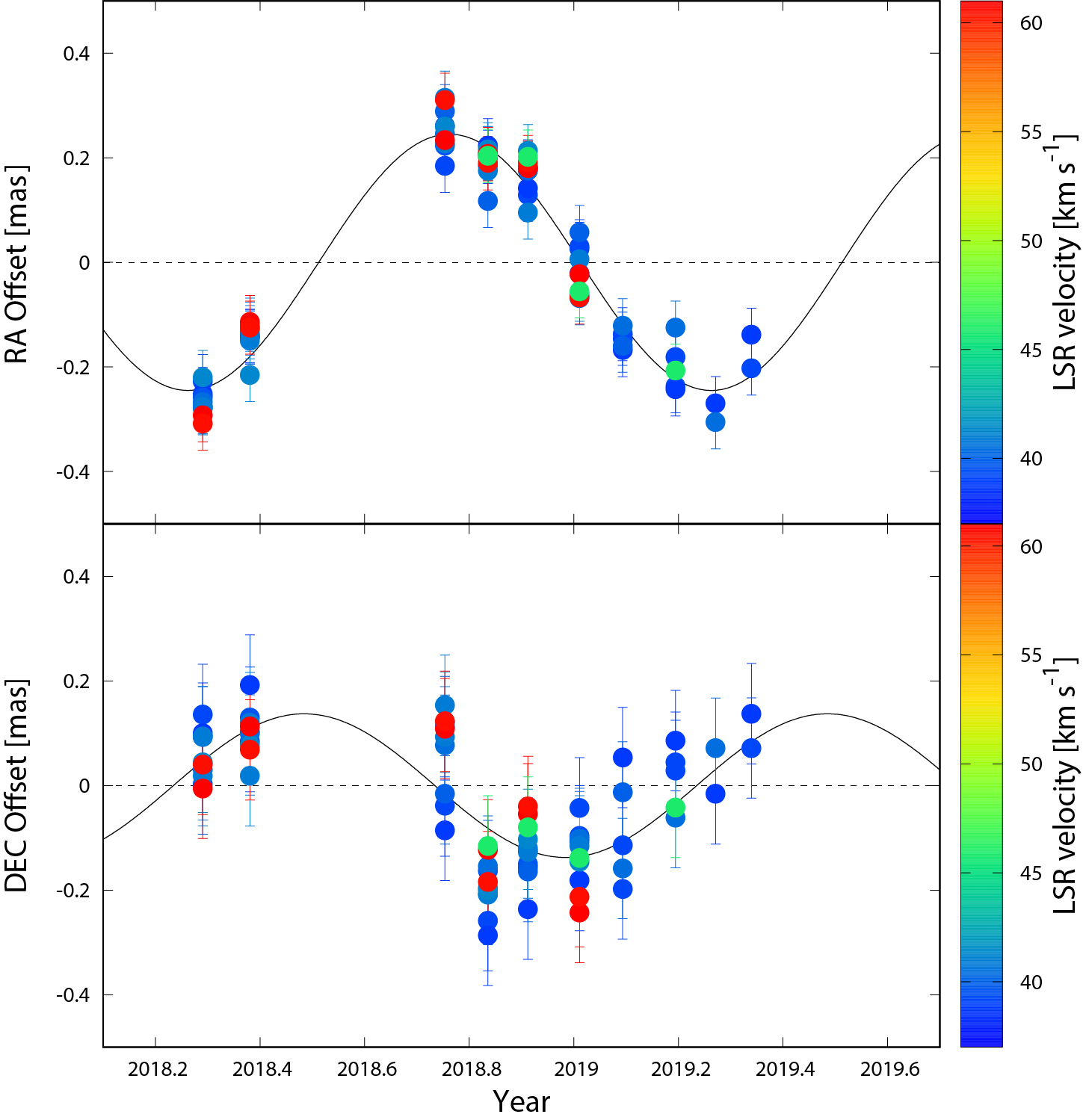}
  \caption{
  The annual parallax of NSV17351 along RA (top) and DEC (bottom) \citep{nak23a}. 
  Results from ten observations are shown with filled circles. 
  The grey scale indicates the LSR velocity $V_{\mathrm{LSR}}$ of each maser spot. 
  The solid curves are the best-fit models obtained from our fit, showing the 0.247 mas parallax. 
}
  \end{center}
\label{fig_parallax}
\end{figure}

To estimate the annual parallax, we track the positions of 22\,GHz H$_2$O maser spots obtained from multiple VLBI images. 
In Figure~\ref{fig-3map} we show examples of maser spot images in the same velocity channel. 
Since the shape of the spot changes gradually with time, we carefully examined the maser structure, its time variation and continuity. 
In this velocity channel, we concluded that the southern components in the maps of Figure~\ref{fig-3map} (b) and (c) are identical to the peak in the map of Figure~\ref{fig-3map} (a).  
Using the 2018-2019 VERA observations of the H$_2$O maser at 22\,GHz, we derived a parallax of 0.247$\pm$0.035 mas for NSV17351, corresponding to a distance of 4.05$\pm$0.59 kpc. 
Figure~\ref{fig_parallax} shows the position offsets after removal of proper motions and the fitted parallax along the RA (top) axis and DEC (bottom) axes. 
The observed data are indicated as filled circles, with their grey scales representing the local standard of rest (LSR) velocities of each maser spot. 
Error bars are 0.05 mas and 0.09 mas in RA and DEC, respectively. 
The solid curves are the best-fit models of the parallax. 
A systemic proper motion of ($\mu_{\alpha}\cos{\delta}, \mu_{\delta}$) $=$ ($-$1.19 $\pm$ 0.11, 1.30 $\pm$ 0.19) mas\,yr$^{-1}$ was also obtained.  

\subsubsection{H$_2$O maser distribution of NSV17351}
The circumstellar distribution and motions of H$_2$O masers in an 80 $\times$ 120 au region around the OH/IR star NSV17351 are shown in the left panel of Figure~\ref{fig_internal}. 
Since the angular proper motions of each maser spot obtained from our VLBI observations are with respect to a position reference source, we have to convert them to motions on the frame fixed to NSV17351. 
To perform this conversion, we take the average motion of the whole maser spots and then subtract it from the original proper motions of each maser spot. 
The detailed procedure has been presented in \cite{nak14}. 
The estimated stellar position of NSV17351 is indicated by a cross symbol whose length represents the position error. 
We can see that the maser spots at different radial velocities are moving outwards from the expected position of the central star. 
As an average, we have derived a three-dimensional outward expansion velocity of the H$_2$O masers of 15.7$\pm$3.3 km\,s$^{-1}$. 

We can see that the bluest maser spots overlap the estimated position of the central star.
In the case of OH masers in OH/IR stars, it is known that the most blue- and red-shifted maser spots are seen at the same position of the central star along the plane of the sky. 
For example, \citet{oro17} revealed that the blue- and red-shifted OH masers coincide with the same position where the central star is assumed to exist. 
Thus, as suggested by the left panel of Figure~\ref{fig_internal}, we can interpret that the most blue-shifted maser spots are superimposed on the position of the central star of NSV17351 along the line of sight. 
They can possibly be explained by the emission being excited along the line of sight to the central star. 

\subsubsection{Position of NSV17351 in our galaxy}
\begin{figure}
\begin{center}
\includegraphics[width=130mm]{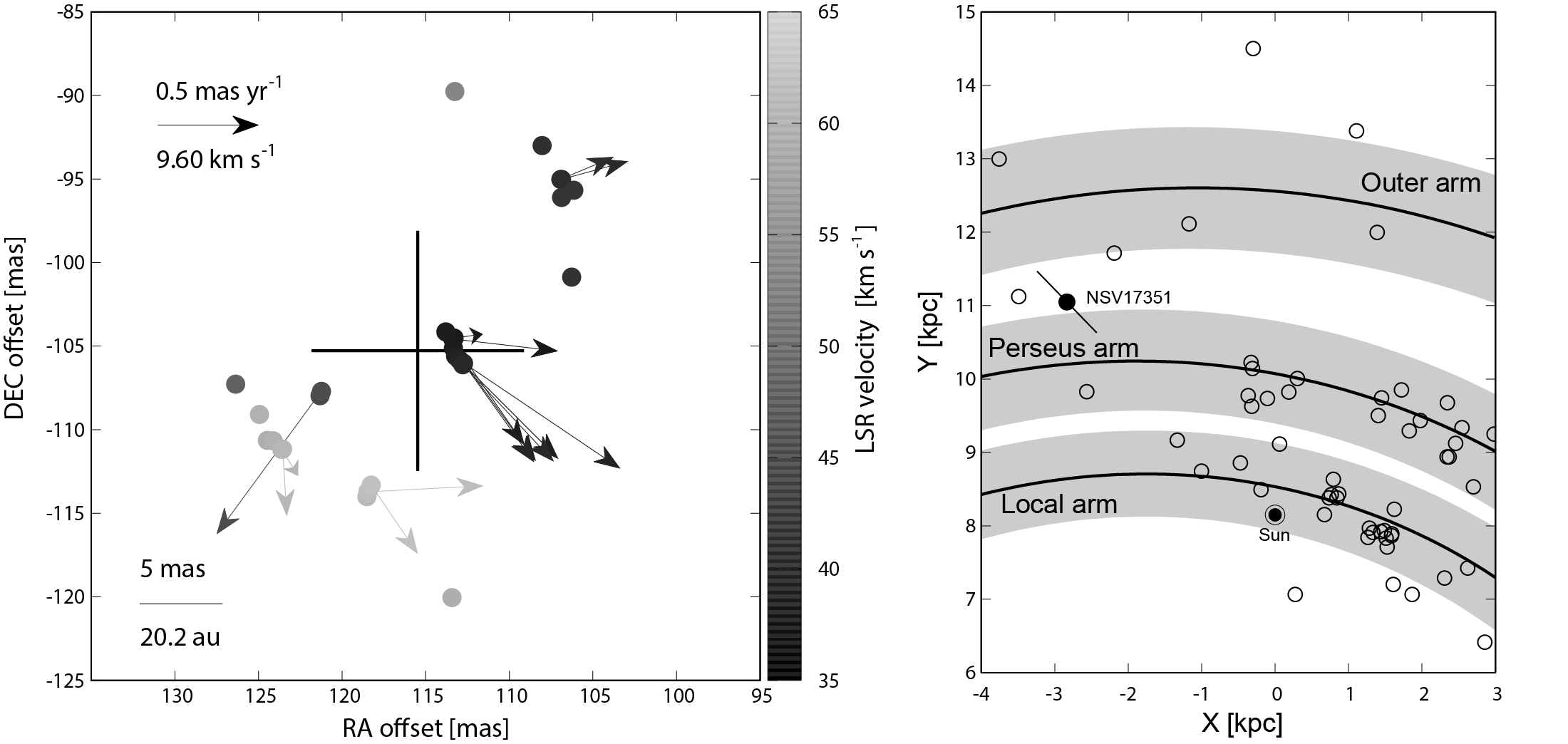}
\end{center}
\caption{
(Left): sky plane distribution and expansion motions of the H$_2$O maser spots of NSV17351 \citep{nak23a}. 
Filled circles indicate maser spots and arrows indicate their internal motions.
A cross symbol indicates the estimated position of the central star. 
(Right): Position of NSV17351 on the face-on view of the Milky Way. 
The Galactic center is at (0, 0) kpc and the Sun is indicated by the symbol ($\odot$) at (0, 8.15) kpc. 
The filled circle with an error bar indicates the position of NSV17351. 
Open circles indicate maser sources with Galactocentric distances $>$ 7 kpc. 
Solid lines and grey regions indicate the centers of three spiral arms and their widths, respectively, reproduced from a study by \citet{rei19}. 
}
\label{fig_internal}
\end{figure}

From the systemic proper motion of NSV17351 derived from our VLBI observations, we can estimate a motion of the source in Galactocentric coordinates. 
In \cite{nak23a}, we derived a three-dimensional position of NSV17351 as 
($X, Y, Z$) $=$ ($-$2.83$\pm$0.12, 11.05$\pm$0.12, $-$0.09$\pm$0.01) kpc, 
where the origin of the coordinate system is the Galactic center. 
We confirm that the $Z$ value of NSV17351, $Z = -0.09 \pm 0.01$ kpc, is within the $Z$ range of SFRs (i.e., $-0.12 < Z < 0.11$ kpc). 
In the right panel of Figure~\ref{fig_internal} we have shown the position of NSV17351 using a filled circle. 
NSV17351 is located a little bit outside of the Perseus arm. 
The location of NSV17351 can be understood by considering its age. 
\citet{fea09} reported that Mira-type variables with a period of 1000 days have an initial mass of 3 to 4\,$M_{\odot}$. 
Assuming this mass, we can estimate the age of NSV17351 to be 1.6$\times$10$^8$ to 3.5$\times$10$^8$ years. 
The age of NSV17351 can be two orders of magnitude greater than the typical age of high-mass SFRs associated with spiral arms, and we can say that we are observing a state in which NSV17351 is leaving the arm in which it was born, but is not yet sufficiently dynamically relaxed. 
With more samples of OH/IR stars representing very long pulsation periods, we can provide observational results for studies of the Galactic spiral arms \citep{nak23a}. 

\subsubsection{OH39.7$+$1.5, IRC$-$30363 (OH/IR star), and AW~Tau (Mira)}
\label{subsec-3src}
In addition to NSV17351, we have observed H$_2$O masers in two OH/IR stars, OH39.7$+$1.5 and IRC$-$30363, and a Mira-type variable, AW~Tau, at 22\,GHz. 

For OH39.7$+$1.5, using two maser spots with radial velocities of 34.6 and 8.6 km\,s$^{-1}$, a preliminary parallax of 0.54$\pm$0.03 mas was obtained. 
This corresponds to a distance of 1.85$\pm$0.10 kpc. 
This parallax measurement is the only one available, as DR3 has no data for OH39.7$+$1.5. 
The averaged proper motion of the two maser spots is 
$(\mu_{\alpha}\cos{\delta}, \mu_{\delta}) = (-0.22\pm0.13, -1.53\pm0.13)$ mas\,yr$^{-1}$. 

For IRC$-$30363, using a maser spot with a radial velocity of 9.72 km\,s$^{-1}$, a parallax of 0.562$\pm$0.201 mas was obtained from our VLBI observations. 
This gives a distance of 1.78$\pm$0.73 kpc. 
In DR3, the parallax of IRC$-$30363 is reported to be 0.241$\pm$0.130 mas. 
They agree within their margins of error. 

For the Mira-type variable AW~Tau, a parallax of 0.449$\pm$0.032 mas was obtained using a maser spot at a radial velocity of $-$9.54 km\,s$^{-1}$. 
This corresponds to a distance of 2.23$\pm$0.16 kpc. 
In DR3, the parallax of AW~Tau is reported to be 0.434$\pm$0.113 mas. 
For this source, the two measurements from the VLBI and DR3 are in very good agreement. 

The pulsation periods of OH39.7$+$1.5, IRC$-$30363 (OH/IR star), and AW~Tau are 1260, 720, and 672 days, respectively. 
All sources show longer pulsation periods than typical Mira-type variables, e.g., lack of Mira-type variables with periods of about 500 days or longer are reported in \citet{hab96}. 
We are currently working on a more detailed analysis of the sources with very long pulsation periods. 

\subsection{Finding a new period-magnitude relation for OH/IR stars in mid-infrared}
Based on the parallax measurements presented in Section~\ref{subsec-3src}, we estimated the absolute K-band magnitudes ($M_\mathrm{K}$) of three AGB stars, AW~Tau, IRC$-$30363, and OH39.7$+$1.5. 
If we compare the three $M_\mathrm{K}$ values with a period-$M_\mathrm{K}$ diagram of the Galactic LPVs, we find that the $M_\mathrm{K}$ value of OH39.7$+$1.5 is far below the expected one. 
About 6 years ago, we investigated the same problem. 
Using published literature results, we compiled $\sim$20 OH/IR stars with very long periods of more than 1000 days. 
The distances of some of the sources were determined using the ``phase-lag method'' (e.g., \citealp{eng15, eto18}). 
For the other sources for which no distance estimate is available, \cite{nak18} derived kinematic distances from their radial velocities. 
Using these distances, the K-band apparent magnitudes were converted to absolute magnitudes ($M_\mathrm{K}$) and presented on a period-$M_\mathrm{K}$ diagram in Figure~1(b) of \cite{nak18}. 
As a result, the distribution of $M_\mathrm{K}$ values of the OH/IR stars shows large scatter. 
We could not find a clear relationship between periods and $M_\mathrm{K}$ values. 
The OH/IR stars are thought to be surrounded by thick layers of circumstellar dust. 
The activity of the central star also affects the outer layers. 
In addition, the spatial structure of the dust layers surrounding the central star is unstable and anisotropic. 
We think that this is one reason for the large scatter seen in the K-band magnitudes. 
In the case of the recent result obtained for the OH/IR star OH39.7$+$1.5, if we assume a thick dust layer or a high mass loss rate, we can possibly explain this darkening by circumstellar absorption. 

At longer wavelengths, re-radiation from the dust shell is known to be dominant. 
To minimise the circumstellar extinction, we have estimated the absolute magnitudes $M_\mathrm{W3}$ in the mid-infrared using the W3 band data from the Wide-field Infrared Survey Explorer (WISE). 
The central wavelength of the WISE W3 band is 12 $\mu$m \citep{wri10}. 
In Table~\ref{table_w3} we have compiled 36 Galactic LPVs whose parallaxes have been determined by VLBI observations. 
Some red giants and SR variables are also included in the table. 
Sources with pulsation periods shorter than 200 days have been excluded. 
Using the parallaxes from VLBI observations $\Pi_{\mathrm{VLBI}}$ in the third column, we derived the absolute magnitude in the W3 band ($M_\mathrm{W3}$) and present it in the seventh column. 
In this table, $\sigma M_{\mathrm{W3}}$ is the error of $M_{\mathrm{W3}}$ estimated by taking into account the parallax error ($\sigma\Pi_{\mathrm{VLBI}}$) and the WISE measurement error. 
The pulsation period $P$ and its logarithm $\log P$ are also given. 
The $M_\mathrm{W3}$ values of the latest VLBI observation sources AW~Tau, IRC$-$30363, and OH39.7$+$1.5 were found to be $-10.71\pm0.21$, $-12.35\pm1.21$, and $-12.90\pm0.40$, respectively. 

\begin{figure}
\begin{center}
\includegraphics[width=130mm]{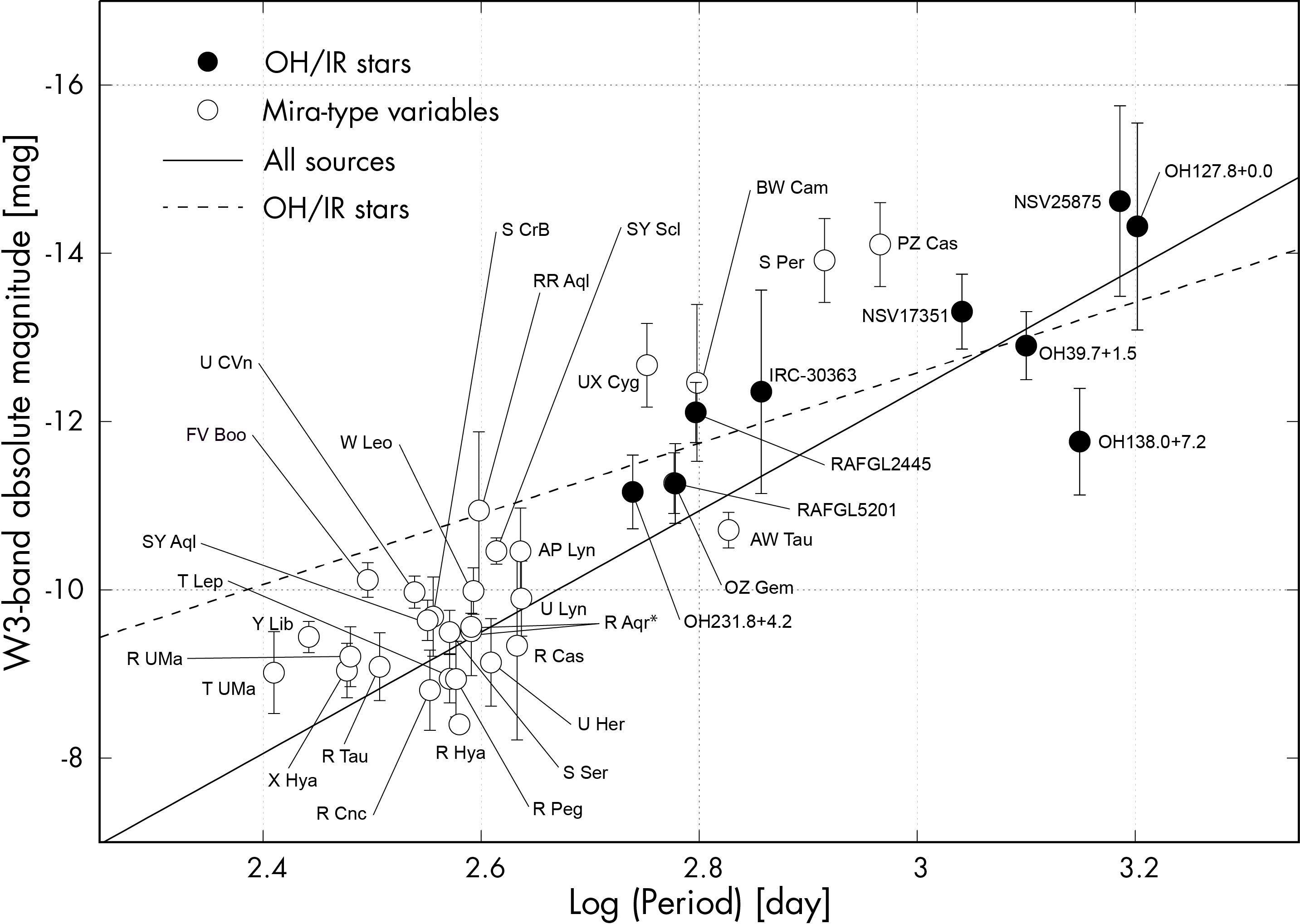}
\end{center}
\caption{
Absolute magnitudes of the WISE W3 band estimated from VLBI parallaxes. 
Filled circles and open circles represent OH/IR stars and Mira-type variables respectively. 
Two lines represent period-$M_{\mathrm{W3}}$ relations in the mid-infrared.
The solid line and dotted lines represent the relations of $M_{\mathrm{W3}} = (-7.21\pm1.18)\log P + (9.25\pm3.09)$ and $M_{\mathrm{W3}} = (-4.20\pm1.33)\log P + (0.01\pm3.89)$, respectively.  
}
\label{fig_plr}
\end{figure}

In Figure~\ref{fig_plr}, we present the pulsation periods and the W3-band absolute magnitudes ($M_\mathrm{W3}$) of the sources in Table~\ref{table_w3}. 
The Mira-type and SR variables are presented with open circles. 
The OH/IR stars are represented by filled circles. 
First, all 37 sources were used to derive the period-$M_\mathrm{W3}$ relation. 
The derived relation is 
\begin{eqnarray}
\label{eq_plr1}
M_{\mathrm{W3}} = (-7.21\pm1.18)\log P + (9.25\pm3.09), 
\end{eqnarray}
which is indicated by a solid line in the figure.  
We then derived another relationship using only the 9 OH/IR stars, represented by filled circles, and obtained a relation of the form 
\begin{eqnarray}
\label{eq_plr2}
M_{\mathrm{W3}} = (-4.20\pm1.33)\log P + (0.01\pm3.89), 
\end{eqnarray}
which is shown with a dotted line. 
At this stage, it is difficult to say that there is an obvious period-$M_\mathrm{W3}$ relation. 
However, if a clear relationship is confirmed, it can be used as a new distance estimator for the AGB sources along the Galactic plane, or for the sources deeply obscured by circumstellar dust. 
It is known that there is a deep silicate absorption feature at $\lambda \simeq$ 8 to 18 $\mu$m in the spectral energy distribution of AGB stars. 
So, ideally, it would be necessary to use bolometric absolute magnitudes $M_\mathrm{bol}$ to avoid the effects of absorption. 
We have tried to do this, but the variance of the $M_\mathrm{bol}$ values is greater than that in the W3 band, and so far our attempt has not been successful. 
For a more detailed investigation of the period-$M_\mathrm{W3}$ relation, we are continuing astrometric VLBI observations of OH/IR stars to increase the number of sources covering a wide period range. 

\begin{table}[th!]
\caption{WISE W3 band absolute magnitude}
\label{table_w3}
\begin{center}
\begin{tabular}{lccccccc} 
    \midrule
Source & Var. & $\Pi_{\mathrm{VLBI}}$ & $\sigma\Pi_{\mathrm{VLBI}}$ & Period & $\log P$ & $M_{\mathrm{W3}}$ & $\sigma M_{\mathrm{W3}}$ \\
       & type & [mas]    &   [mas]              &  ($P$) [d]    &   &   [mag]      &  [mag]         \\ 
   \midrule
SY Scl & Mira & 0.75 & 0.03 & 411 & 2.61 & $-$10.46 & 0.16 \\
OH127.8$+$0.0 & OH/IR & 0.22$^{\dag}$ & 0.08 & 1591 & 3.20 & $-$14.32 & 1.23 \\
S Per & SRc & 0.413 & 0.017 & 822 & 2.92 & $-$13.92 & 0.50 \\
OH138.0$+$7.2 & OH/IR & 0.52$^{\dag}$ & 0.09 & 1410 & 3.15 & $-$11.76 & 0.63 \\
R Tau & Mira & 2.04 & 0.05 & 321 & 2.51 & $-$9.09 & 0.40 \\
T Lep & Mira & 3.06 & 0.04 & 372 & 2.57 & $-$8.94 & 0.29 \\
BW Cam & Mira & 0.749 & 0.189 & 628 & 2.80 & $-$12.46 & 0.93 \\
AW Tau & Mira & 0.45 & 0.03 & 672 & 2.83 & $-$10.71 & 0.21 \\
RAFGL\,5201 & OH/IR & 0.61$^{\dag}$ & 0.04 & 600 & 2.78 & $-$11.26 & 0.47 \\
AP Lyn & Mira & 2.01 & 0.04 & 433 & 2.64 & $-$10.46 & 0.51 \\
U Lyn & Mira & 1.27 & 0.06 & 434 & 2.64 & $-$9.90 & 0.45 \\
NSV17351 & OHIR & 0.247 & 0.035 & 1100 & 3.04 & $-$13.31 & 0.45 \\
OZ Gem & Mira & 0.806 & 0.039 & 598 & 2.78 & $-$11.27 & 0.36 \\
OH231.8$+$4.2 & OH/IR & 0.61$^{\dag}$ & 0.03 & 548 & 2.74 & $-$11.16 & 0.44 \\
R Cnc & Mira & 3.84 & 0.29 & 357 & 2.55 & $-$8.81 & 0.48 \\
X Hya & Mira & 2.07 & 0.05 & 300 & 2.48 & $-$9.04 & 0.32 \\
R UMa & Mira & 1.97 & 0.05 & 302 & 2.48 & $-$9.21 & 0.36 \\
W Leo & Mira & 1.03 & 0.02 & 392 & 2.59 & $-$9.99 & 0.28 \\
T UMa & Mira & 0.96 & 0.15 & 257 & 2.41 & $-$9.02 & 0.49 \\
U CVn & Mira & 0.911 & 0.031 & 346 & 2.54 & $-$9.97 & 0.19 \\
R Hya & Mira & 7.93 & 0.18 & 380 & 2.58 & $-$8.40 & 0.07 \\
FV Boo & Mira & 0.97 & 0.06 & 313 & 2.50 & $-$10.12 & 0.21 \\
Y Lib & Mira & 0.855 & 0.050 & 277 & 2.44 & $-$9.44 & 0.19 \\
S CrB & Mira & 2.39 & 0.17 & 360 & 2.56 & $-$9.68 & 0.47 \\
S Ser & Mira & 1.25 & 0.04 & 372 & 2.57 & $-$9.50 & 0.26 \\
U Her & Mira & 3.76 & 0.27 & 406 & 2.61 & $-$9.14 & 0.52 \\
IRC$-$30363 & OH/IR & 0.56$^{\dag}$ & 0.2 & 720 & 2.86 & $-$12.35 & 1.21 \\
OH39.7$+$1.5 & OH/IR & 0.55$^{\dag}$ & 0.03 & 1259 & 3.10 & $-$12.90 & 0.40 \\
RAFGL\,2445 & OH/IR & 0.64$^{\dag}$ & 0.01 & 626 & 2.80 & $-$12.11 & 0.36 \\
RR Aql & Mira & 1.58 & 0.4 & 396 & 2.60 & $-$10.94 & 0.93 \\
SY Aql & Mira & 1.10 & 0.07 & 356 & 2.55 & $-$9.64 & 0.24 \\
UX Cyg & Mira & 0.54 & 0.06 & 565 & 2.75 & $-$12.67 & 0.50 \\
NSV\,25875 & OH/IR & 0.38$^{\dag}$ & 0.13 & 1535 & 3.19 & $-$14.62 & 1.13 \\
R Peg & Mira & 2.76 & 0.28 & 378 & 2.58 & $-$8.94 & 0.45 \\
R Aqr & Mira & 4.7 & 0.8 & 390 & 2.59 & $-$9.51 & 0.53 \\
      & Mira & 4.59 & 0.24 & 390 & 2.59 & $-$9.56 & 0.16 \\
PZ Cas & SRc & 0.356 & 0.026 & 925 & 2.97 & $-$14.10 & 0.50 \\
R Cas & Mira & 5.67 & 1.95 & 430 & 2.63 & $-$9.33 & 1.12 \\
    \midrule
    \end{tabular}
    \tabnote{${\dag}$ The parallax values are preliminary results from 
    our analysis and have not yet been published.  
}
\end{center}  
\end{table}   

\section{Summary}
We have been observing Mira-type variables and OH/IR stars with the astrometric VLBI method. 
The obtained distances and stellar parameters help us to understand the evolutionary relationship of the subclasses in the AGB phases. 
The VERA array was used for all VLBI observations in our study. 
The masers used are the H$_2$O one at 22\,GHz and the SiO one at 43\,GHz. 
The phase referencing technique was used to accurately measure the distances of stars ranging from a few hundred pc to several kpc. 
Due to the properties of the circumstellar matter and its time variability, the parallax measurements of stars surrounded by thick dust shells in optical bands is sometimes very difficult. 
This can sometimes make parallax measurements with {\em Gaia} difficult.
The VLBI and {\em Gaia} are complementary for distance measurements of AGB stars. 

The results for NSV17351, AW~Tau, IRC$-$30363, and OH39.7$+$1.5 are presented.
The pulsation period of NSV17351 was determined from the time variability of the H$_2$O maser. 
Parallax, circumstellar distribution of the masers and kinematics of NSV17351 were also presented. 
Absolute magnitudes in the near- and mid-infrared bands of OH/IR stars with very long pulsation periods were studied. 
Using the WISE W3 band data, a period-magnitude relation in the WISE W3 band $M_{\mathrm{W3}} = (-7.21\pm1.18)\log P + (9.25\pm3.09)$ was found for the Galactic AGB stars. 
For further understanding, we need more detailed measurements of the circumstellar masers of AGB stars of different types, pulsation periods, and masses. 

\section*{Discussion}

\textbf{Question (Whitelock)}: 
These VLBI measurements are very important for deriving distances for the stars that {\em Gaia} can’t reach. 
Most of their luminosity will be in the dust and the amplitudes, even at long wavelengths, are very large, several magnitudes. 
You can only observe the O-rich AGB stars with H$_2$O and SiO Masers. 
Metal-weak stars are C-rich and will not have these lines, so I wonder if you can do anything with the CO line from the C-rich stars.

\textbf{Answer}: 
Thank you for your comment, Patricia. 
One thing that I wanted to share at this conference was the importance of VLBI measurements of the parallaxes of dusty AGB stars, and I think we did that to some extent. 
The CO lines are usually observed in dusty AGB stars. 
The existence of CO masers has also been reported by \cite{vle21}. 
However, at present I think that astrometric VLBI observations of the CO maser are difficult due to the accuracy of the absolute position and the sensitivity. 
So I do not expect to be able to do the same studies with the CO masers any time soon. 
Regarding the CO line, if we can somehow select OH/IR stars with very long pulsation periods, they could be massive and young, and they are expected to show CO line emission. 
Since observing the CO line can give a better estimate of radial velocities than maser emissions, we can estimate their distances using the kinematic distance method and use them for the PL relation or to study the Galactic dynamics.

\textbf{Question (Jiang)}: 
It’s very nice to see that the maser observation can be more accurate than {\em Gaia} for the nearby dust-obscured stars. 
I just wonder how the other astrometric parameters, proper motions, measured using maser compare to {\em Gaia}?

\textbf{Answer}: 
Thank you for your comment, Biwei. 
VLBI measures the sky plane motion of each maser spot, which is distributed around the central AGB star. 
The maser spots show outward motion with respect to the central star. 
Thus, the proper motions of each maser spot obtained from VLBI observations are the sum of (1) the systemic proper motion of the central star and (2) the outward motion of the maser with respect to the central star.
We average all the proper motions of the detected maser spots to estimate the systemic proper motion of the central star, which {\em Gaia} can measure directly.
The difference in proper motions between {\em Gaia} and VLBI gives us information about the outward motion of the maser spots with respect to the central star.
In \cite{nak14}, we presented the procedure to derive the systemic proper motion of the star from our VLBI measurements.


\bibliographystyle{mnras}
\bibliography{main}

\end{document}